\documentclass[letterpaper]{article}
\usepackage{amssymb}
\usepackage{amsmath,enumerate}
\usepackage{graphicx}
\usepackage{mathpazo}
\usepackage[mathpazo]{flexisym}
\usepackage{breqn}
\begin{document}

\title{Formal Verification of Safety Properties for Ownership Authentication Transfer Protocol}
\author{Swaraj Bhat,~Pradeep B.H,~Keerthi S.Shetty  and Sanjay Singh\thanks {Sanjay Singh is with the Department of Information and Communication Technology, Manipal Institute of Technology, Manipal University, Manipal-576104, INDIA, E-mail: sanjay.singh@manipal.edu}}

\maketitle
\begin{abstract}
In ubiquitous computing devices, users tend to store some valuable information in their device. Even though the device can be borrowed by the other user temporarily, it is not safe for any user to borrow or lend the device as it may cause private data of the user to be public. To safeguard the user data and also to preserve user privacy we propose and model the technique of ownership authentication transfer. The user who is willing to sell the device has to transfer the ownership of the device under sale. Once the device is sold and the ownership has been transferred, the old owner will not be able to use that device at any cost. Either of the users will not be able to use the device if the process of ownership has not been carried out properly. This also takes care of the scenario when the device has been stolen or lost, avoiding the impersonation attack. The aim of this paper is to model basic process of proposed ownership authentication transfer protocol and check its safety properties by representing it using CSP and model checking approach. For model checking we have used a symbolic model checker tool called NuSMV. The safety properties of ownership transfer protocol has been modeled in terms of CTL specification and it is observed that the system satisfies all the protocol constraint and is safe to be deployed.
\end{abstract}

\section{Introduction}
A ubiquitous computing (Ubicomp) or pervasive computing environment is imagined as a system with numerous invisible computers, sensors and actuators interacting with the user devices such as PDAs, Laptops, Mobile Phones etc. Data about the individuals who are a part of the ubiquitous environment is constantly being generated, transmitted, manipulated and stored. The user data present in the environment (in device or servers) is very sensitive. Protecting private data of every user in the environment is a major concern. Also in the this era of the mobile environment the user owns more than one portable devices like the PDAs, Laptops, Mobile Phones etc. with varying computing capabilities in order to access the variety of services that are being provided by the service providers. At times the user may tend to sell the device he owns. Since the device consists of the valuable information of the user and also will have the access to the valuable information present at the server, care should be taken to delete the information of the previous owner and store the details of the new owner in the device as well as the server.
\par
Paulo Tam and Jan Newmarch \cite{1} in their work have suggested the concept of transferring the ownership of the device. The owner (old owner) of the device will send the message to the device itself that it is being bought by the other user (new owner). The device will send the message to the new owner saying that its ownership is about to change to you (new user), do you accept or reject. The new owner sends the response to the device and the object will in turn send an acknowledgment on the status of the transfer to the old owner. However when the owner of the device is selling the device to the new owner, sending the message to the device itself does not seem feasible. Moreover to which device of the user, the device under sale is sending the message is not known. It is feasible if the new owner of the device has one more device under his ownership. But if the user has no other device previously and it is his first device then there is no possibility for the device under sale to send the message to its new owner asking his consent on the ownership transfer. In ubiquitous environment the ownership transfer has to be informed to the central server instead of informing to the device under sale. 
\par
Jurgen Bohn \cite{2} has mentioned that the user can borrow or lend the device to his friend or the stranger. The data of a particular user can be retrieved from the personalization server at any time and from anywhere for a specific time. Once the time limit is exceeded, the session will expire and the user needs to quit the session or restart it. After using the device, the user can release the device and return it back to the owner of the device. But the very basic idea of sharing the personal device with a friend or a stranger may cause information to be public. This could be due to the other user being malicious (intentionally causing harm) by installing some kind of software which can record the data of the user or simply careless (unintentionally installing malicious software which can access the users data). Due attention should be paid to the fact that the device could come with old data, if the transfer is incomplete due to technical reasons such as network congestion or lack of connectivity. The owner of the device may also turn out to be malicious with respect to the other user. The user may install a software that records all the data that has been retrieved and sent from that device before encryption and after decryption. Later the user may be subjected to the impersonation attack. Moreover when the time limit is exceeded, there are chances that the user may have to end the session or restart it due to network latencies or unresponsive server when the user is trying to retrieve or release the data.
\par
Yongming Jin et al \cite{3} has described the transfer of RFID from the old owner to the new owner. They define a protocol to safeguard the privacy of the respective owners by putting the clean stop between the transactions of the old and the new owners by means of a secret. The authors have suggested the use of RFIDs for the ownership transfer. But there are many security concerns with respect to the RFID tags. One of the primary RFID security concern is the illicit tracking of RFID tags. The tags can be read by anyone in the world and if the person who read the tag is malicious can pose a risk by either impersonating the user or trying to manipulate the user data and use it for a wrong purpose. RFIDs working at a shorter range are vulnerable to skimming and eavesdropping. Even though certain RFID tags use cryptographic features, the cost and power requirements are very high when compare to the simpler RFID tags. Thus, the cost and power limitation has compelled some manufacturers to implement cryptographic tags using substantially weak encryption schemes, which are weak to resist the sophisticated attack. Moreover, the power available in the handheld devices is limited; these tags cannot be incorporated in the devices.
\par
Abdullah M. Alaraj \cite{4} in his paper suggest that the users have to go to some officially designated place for buying or selling the merchandise and to complete the process of ownership transfer. He also makes an assumption that the certain equipments are required for ownership transfer and tries to improve the fairness by including the transfer of money through the bank servers. However going to an officially designated place that deals with buying or selling of merchandise is suitable only to the goods like cars or for real estate. This scenario will not be suitable when applied to ubiquitous computing devices. The process of ownership transfer requires only a Central Key Server(CKS) and a device meant for sale. Submitting users bank details to the third party might be risky at the time of payment. Even thought if the system provides the best servers for transaction and promotes the users to submit their bank details to the device in an office meant for buying and selling of the merchandise, the device or the system in that office might turn out to be malicious. 

\par
In this paper we have modeled a newly proposed ownership authentication transfer protocol \cite{pra}\cite{ps} which overcomes the limitations of the existing ownership transfer protocol. To the best of our knowledge any form of work in the field of formalizing ownership authentication transfer in ubiquitous computing devices has been minimal. In this paper, we have described basic process of ownership authentication transfer and formalized the safety properties using CSP approach. The safety properties of the proposed protocol is verified using a symbolic model checking approach. We have used a tool called New Symbolic Model Verifier (NuSMV) \cite{7}. It searches the entire possible state space and checks for the correctness of the various specifications.

\par
The remaining paper is organized as follows. Section \ref{a} explains the operation of the proposed ownership authentication transfer protocol. Section \ref{b} discusses modeling of the proposed protocol using CSP approach. Section \ref{c} and \ref{d} briefly discusses the concept of model checking and model checker tool used. Section \ref{e} explains the modeling about of the safety properties in NuSMV. Section \ref{f} discusses about the model checking results obtained for the safety properties for the proposed protocol. Finally, a conclusion has been drawn in section \ref{g}.

\section{Device Ownership Authentication Transfer Protocol}
\label{a}
In this paper we propose a secure and fair protocol for ownership transfer of the ubiquitous computing devices. The user who is buying an old device from the other user has to undergo this process in order to successfully acquire the ownership of the device and start using it in the ubiquitous environment.\\

\textbf{Assumption:} The value or the price of the device has been agreed upon between the users before transferring the ownership of the device.\\

\textbf{Requisite:} The users should be in physical proximity and the whole process has to be carried out in the device which is under sale.

\begin{table*}[bpht!]
\centering
	\caption{Notations Used} 
	\label{tab}
	\begin{tabular}{|c|p{2in}|c|p{1.5in}|}
	\hline 
	\textbf{Symbol}&\textbf{Meaning}&\textbf{Symbol}&\textbf{Meaning} \\ \hline 
	
	$E_{P_{CKS}}$&Encryption Using Public key of CKS&$CKS$&Central Key Server\\ \hline 
			$N_A$&nonce generated by A& $N_B$&Nonce generated by B \\ \hline
			$ID_A$&User ID or User Name of the user A&$ID_B$&User ID or User Name of the user B\\ \hline 
			$P_{CKS}$&Public key of the CKS&$Ack$& Acknowledgment\\ \hline
			$PW_A$&Password of the User A&$Temp ID$& Temporary ID or Pseudo ID\\ \hline 
			$OTC$&Ownership Transfer Confirmation&$OTR$&Ownership Transfer Request\\ \hline
		  
	\end{tabular}
\end{table*}	 
 
\begin{figure}[bpht!]
\centering
\includegraphics[width=12cm,height=8cm]{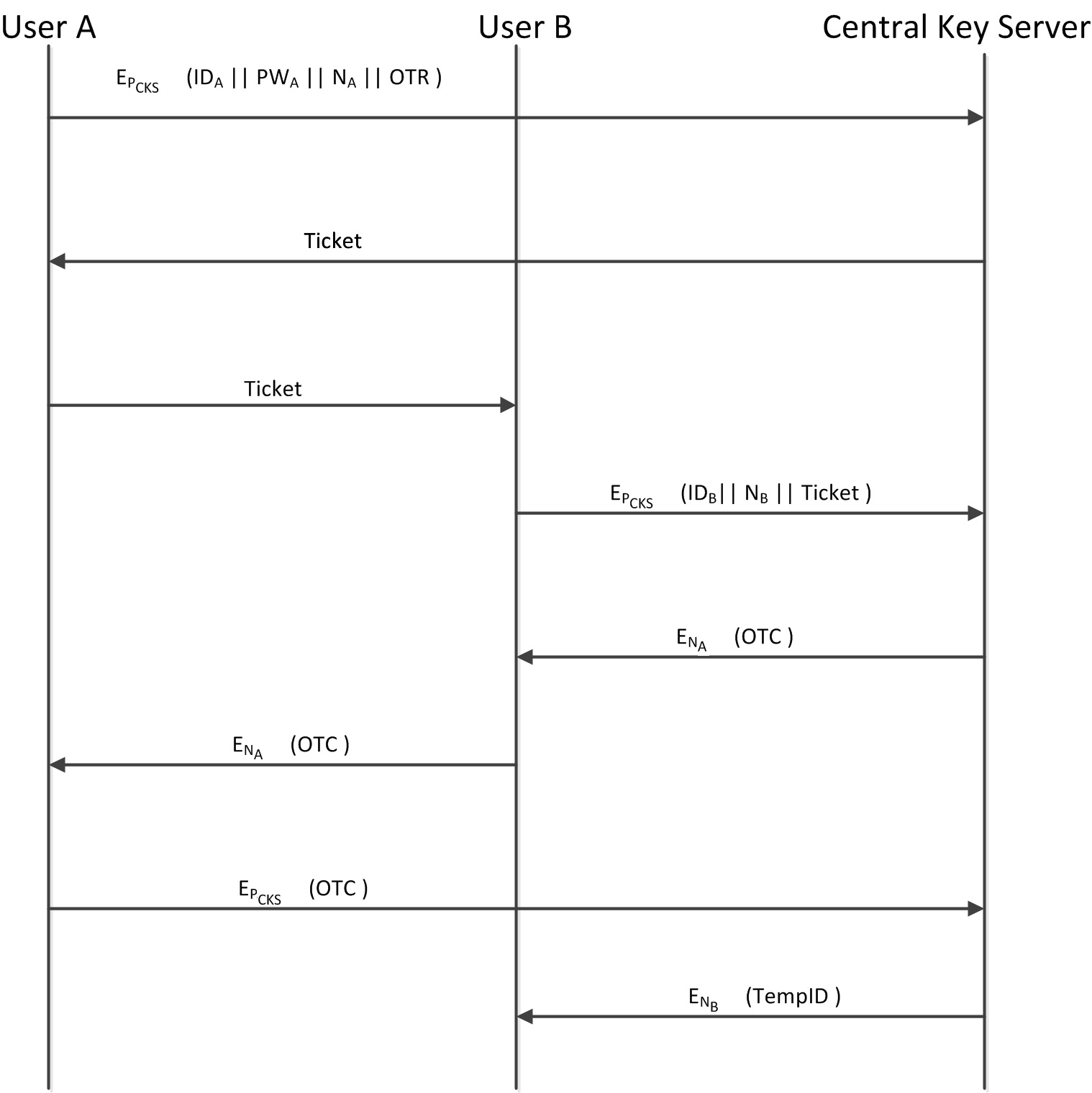}
\caption{Diagram Showing Device Ownership Transfer Process}
\label{fig:1}
\end{figure}

The previously existing user should introduce the new owner of the device to the CKS, in other words user A must transfer the ownership authentication credentials to the new user B. Once the new owner is introduced, the CKS will delete the credentials of the previous owner and save the credentials of the new owner for the same device. Once the ownership rights has been transfered to the new user, the old user at any cost will not be able to use the device. If in case the whole process of ownership transfer as mentioned below is not completed, neither of the users will be able to use the device. This also takes care of the scenario that if a device is stolen, the thief cannot use the device. The proposed ownership transfer protocol for a given ubiquitous device is explained below.
\begin{enumerate}
				\item $ U_A\rightarrow CKS \ :\ E_{P_{CKS}} (ID_{A} \| PW_A \| N_A\| OTR)$ \\ 
				The user A (Old User) sends the message to the CKS. The message consists of the user A ID, password of the user A, nonce of the user A and Ownership Transfer Request (OTR). This message is encrypted using the public key of the CKS. OTR consists of the ID of the user selling the device, ID and nonce of the user buying the device. OTR is also encrypted using the public key of the CKS, where $OTR = E_{P_{CKS}} (ID_A \| ID_B \| N_B)$. In this step the user A will introduce user B to the CKS.  
				\item  $CKS \rightarrow U_A \ :\ Ticket$\\ 
				In response to the user A's request for ownership transfer, the CKS sends a ticket to the user A. The Ticket consists of the acknowledgment for ownership transfer to the user B. The ticket is encrypted using the public key of the CKS.
				\item $U_B \rightarrow CKS \ :\ E_{P_{CKS}}(ID_{B} \| Ticket \| N_B)$\\ 
				The user A will now hand over the device to the user B. Now the user B sends his credentials to the CKS. The user needs to send user ID, nonce and the ticket got by the user A. The ticket will be in the device itself.   
			 \item $CKS \rightarrow U_B \ :\ E_{N_{A}}(\mbox{OTC})$\\ 
					Once the CKS receives the credentials of User B, the CKS sends the Ownership Transfer Confirmation(OTC) to the user B by encrypting it using nonce of the user A. This message consists of the information about the money to be transferred and the account details of the destination account. 
				\item $U_A \rightarrow CKS \ :\  E_{P_{CKS}}(\mbox{OTC})$\\ 
				The user B will hand over the device to the user A and the user A will decrypt the message, read the acknowledgment and then he sends the acknowledgment back to the CKS by encrypting it using the public key of CKS. By sending the acknowledgment back to the CKS, he confirms the ownership transfer of the device. Signing a particular message twice is required to strike the fairness in the deal. There may be some chances where either of the users may turn to be malicious. This is done in order to obtain a confirmation from the user who is selling the device. 
				\item $CKS \rightarrow U_B \ :\ E_{N_{B}}(TempID)$\\ 
				On receiving the message, CKS completes the ownership transfer of the device by sending the temp ID  to the user B. The temp ID is encrypted using the nonce of the user B. 
								
			\end{enumerate}
			
				The above explained process of device ownership transfer is summarized in the Fig. \ref{fig:1}.

\section{Modeling Device Ownership Authentication Transfer Protocol using CSP Approach}
\label{b}

Communicating Sequential Processes (CSP) \cite{masp1}\cite{masp2} is a notation for describing systems of parallel agents that communicate by passing messages between them. Security protocols work through the interaction of a number of processes in parallel that send each other messages. The typical security protocol involves several agents (often two: an initiator and a responder) and perhaps a server that performs some service such as key generation, translation or certification. 

The Yahalom Protocol \cite{yp} representation of Ownership Authentication Transfer Protocol is as follows.
\begin{enumerate}[M1]
\item A $\rightarrow CKS  \ :\  \{ID_{A} \cdot PW_A \cdot  N_{A} \cdot OTR \}_{P_{CKS}}$ 
\item CKS $\rightarrow A  \ :\   Ticket $
\item B $\rightarrow CKS  \ :\  \{ID_{B} \cdot ticket \cdot  N_{B}\}_{P_{CKS}}$ 
\item CKS $\rightarrow  B  \ :\  \{OTC\}_{N_{A}}$ 
\item A $\rightarrow CKS  \ :\  \{OTC\}_{P_{CKS}}$ 
\item CKS $\rightarrow B  \ :\  \{TempID\}_{N_{B}}$ 
\end{enumerate}

The basic process of Ownership Authentication Transfer Protocol involves two agents: old owner and new owner of the device. The CSP description of protocol is given by equation \eqref{eq:1}.
\begin{figure*}[bpht!]
\begin{dmath}\label{eq:1}
OldOwner(A,N_{A})=env?B:NewOwner \rightarrow send \cdot A \cdot CKS \cdot \{ID_{A}\cdot PW_{A} \cdot N_{A} \cdot OTR\}_{P_{CKS}} \cdot m \rightarrow \\
\begin{array}{c} \square \\N_{A} \in Nonce \\ B \in NewOwner \\ m \in  message \end{array} \left(\begin{array}{c} receive \cdot CKS \cdot A \cdot Ticket \rightarrow \\ send \cdot A \cdot CKS \cdot m \cdot \{OTC\}_{P_{CKS}} \rightarrow \\ 
Session(A,B,P_{CKS},N_{A},N_{B})\end{array} \right)
\end{dmath}
\end{figure*}

The key $P_{CKS}$ is the key of CKS that shares with Old Owner and New Owner of the device. The message is encrypted using public key of CKS. The $env?B\:\ NewOwner$ is a representation how the processes local environment might tell it to open a session with agent B, this is formally expressed by equation \eqref{eq:2}.
\begin{figure*}[bpht!]
\begin{dmath}\label{eq:2}
NewOwner(B,N_{B})= \begin{array}{c} \square \\N_{A}\in Nonce \\ A \in OldOwner \end{array} \left(\begin{array}{c} send \cdot B \cdot CKS \cdot \{ID_{B} \cdot Ticket \cdot N_{B}\}_{P_{CKS}}     \rightarrow \\ receive \cdot CKS \cdot B \cdot \{OTC\}_{N_{A}} \rightarrow \\
receive\cdot CKS \cdot B \cdot  \{TempID\}_{N_{B}} \rightarrow \\ 
Session(B,A,P_{CKS},N_{A},N_{B})\end{array} \right)
\end{dmath}
\end{figure*}

Then Yahalom protocol is described as combination of users and servers. The Yahalom Process is expressed as:\\

$Yahalom = OldOwner | NewOwner | Central Key Server$.\\

When Intruder is present in the environment, the process is expressed as:\\
$System = Yahalom | Intruder$\\

\subsection{Safety Properties}
When Intruder is present in the environment, safety properties will be defined by introducing additional information into protocol descriptions to enable a description of what is expected of the system at particular points during a run of the protocol. The user who is buying an old device from the other user has to take care of safety properties in order to successfully acquire the ownership of the device and start using it in the ubiquitous environment.
\subsubsection{Secrecy}	
The old owner sends his credentials and OTR to the CKS.	This message is kept secret until ownership of the device is transferred to the authenticated new owner. The message \textit{Claim\_Secret} will be inserted at the end of the description of the protocol run by old owner. Intruder cannot obtain any details during a run of the protocol whenever its secrecy is claimed.
\par An event $Claim\_Secret \cdot OldOwner \cdot NewOwner \cdot$ message is used. This says that the message is kept secret during the run of the protocol.

The CSP description of secrecy property is given by equation \eqref{eq:3} and \eqref{eq:4}.
\begin{figure*}[bpht!]
\begin{dmath}\label{eq:3}
OldOwner(A,N_{A})=env?B \ :\ NewOwner \rightarrow send \cdot A \cdot CKS \cdot \{ID_{A}\cdot PW_{A} \cdot N_{A} \cdot OTR\}_{P_{CKS}} \cdot m \rightarrow \\
\begin{array}{c} \square \\N_{A}\in Nonce \\ B \in NewOwner \\ m \in  message \end{array} \left(\begin{array}{c} receive \cdot CKS \cdot A \cdot Ticket \rightarrow \\ send \cdot A \cdot CKS \cdot m \cdot \{OTC\}_{P_{CKS}}\rightarrow \\ 
signal \cdot Claim\_Secret \cdot A \cdot B \cdot N_{B}\rightarrow \\
Session(A,B,P_{CKS},N_{A},N_{B})\end{array} \right)
\end{dmath}
\end{figure*}

\begin{figure*}[bpht!]
\begin{dmath}\label{eq:4}
NewOwner(B,N_{B})= \begin{array}{c} \square \\N_{A}\in Nonce \\ A\in OldOwner \end{array} \left(\begin{array}{c} send \cdot B \cdot CKS \cdot \{ID_{B} \cdot Ticket \cdot N_{B}\}_{P_{CKS}}     \rightarrow \\ receive \cdot CKS \cdot B \cdot \{OTC\}_{N_{A}} \rightarrow \\
receive\cdot CKS \cdot B \cdot  \{TempID\}_{N_{B}} \rightarrow \\ 
Session(B,A,P_{CKS},N_{A},N_{B})\end{array} \right)
\end{dmath}
\end{figure*}
\subsubsection{Authentication}
The old owner who is selling the device, initiates ownership transfer process of the device. In order to formalize authentication properties, two events are introduced during run of the protocol.\\
\begin{itemize}
\item Commit $\cdot$ NewOwner $\cdot$ OldOwner \\ 
This says that NewOwner has completed a protocol run apparently with Old Owner.\\
\item Running $\cdot$ OldOwner $\cdot$ NewOwner \\ 
This says that OldOwner is following a protocol run apparently with NewOwner.\\
\end{itemize}
 	
The CSP description for authentication property of device ownership transfer is given by equation \eqref{eq:5} and \eqref{eq:6}.
\begin{figure*}[bpht!]
\begin{dmath}\label{eq:5}
OldOwner(A,N_{A})=env?B \ :\ NewOwner\rightarrow send \cdot A \cdot CKS \cdot \{ID_{A}\cdot PW_{A} \cdot N_{A} \cdot OTR\}_{P_{CKS}} \cdot m \rightarrow \\
\begin{array}{c} \square \\N_{A}\in Nonce \\ B\in NewOwner \\ m \in  message \end{array} \left(\begin{array}{c} receive \cdot CKS \cdot A \cdot Ticket \rightarrow \\ 
signal \cdot Running\_OldOwner \cdot A \cdot B \cdot N_{A} \cdot N_{B}\rightarrow \\
send \cdot A \cdot CKS \cdot m \cdot \{OTC\}_{P_{CKS}}\rightarrow \\ 
Session(A,B,P_{CKS},N_{A},N_{B})\end{array} \right)
\end{dmath}
\end{figure*}

\begin{figure*}[bpht!] 	
\begin{dmath}\label{eq:6}
NewOwner(B,N_{B})= \begin{array}{c} \square \\N_{A}\in Nonce \\ A\in OldOwner \end{array} \left(\begin{array}{c} send \cdot B \cdot CKS \cdot \{ID_{B} \cdot Ticket \cdot N_{B}\}_{P_{CKS}}     \rightarrow \\ receive \cdot CKS \cdot B \cdot \{OTC\}_{N_{A}} \rightarrow \\
receive\cdot CKS \cdot B \cdot  \{TempID\}_{N_{B}} \rightarrow \\ 
signal \cdot Commit\_NewOwner \cdot A \cdot B \cdot N_{A} \cdot N_{B}\rightarrow \\
Session(B,A,P_{CKS},N_{A},N_{B})\end{array} \right)
\end{dmath}
\end{figure*}

\section{Model Checking}
\label{c}
As systems to be designed become more and more complicated, it is not sufficient at all to check the correctness of designs only by simulation. Subtle design errors can easily survive even under intensive and massive simulation. Also, detecting design errors in the late design stages is extremely costly and must be avoided as much as possible.
\par Formal verification is to mathematically prove that the behavior allowed by given specification (properties) contains
the behavior performed by designs. It is essentially an exhaustive check on every possible behavior of designs that is related to the given specification.
\par Model checking is an automatic method to prove such correctness and is now becoming to be widely used in real design environments. Model checking is basically an exhaustive search in all possible states in the designs by checking whether the given specification is satisfied in all of them. That is mostly an implicit exhaustive search on state space of designs in the sense that state space of designs are represented symbolically instead of individually. This is why state-of-the-art model checking programs can verify designs having up to $10^{100}$ or more states \cite{5}.
\par Specification for model checking is a set of properties that the designs must satisfy. Property can be described either in temporal logic or automaton. Temporal logic is an extension to traditional logic with temporal operators by which we can describe relationships among variables in different time frames. 

\section{NuSMV}
\label{d}
NuSMV is a symbolic model verifier tool for the formal verification of finite state systems. NuSMV allows us to check finite state systems against specifications in the linear temporal logic and Computational Tree Logic (CTL) \cite{6}. The input language of NuSMV is designed to allow the description of finite state systems that range from completely synchronous to completely asynchronous \cite{7}. It provides a modular hierarchical description and reusable component. NuSMV is mostly used for describing the transitions of a finite Kripke Structure \cite{kp}.
\par	NuSMV works only with finite data types such as boolean, scalar and fixed array. The main features of NuSMV are its functionalities, architecture and quality of implementation. NuSMV allows analysis of specification expressed in CTL and Linear Temporal Logic (LTL) using Binary Decision Diagram (BDD) and SAT-based checking \cite{6}.\\
	 We have used NuSMV for modeling basic process of ownership authentication transfer protocol and verification of its safety properties.

\section{Modeling Safety Properties in NuSMV}
\label{e}
\begin{figure*}[bpht!]
	\centering
		\includegraphics[width=8cm,height=6cm]{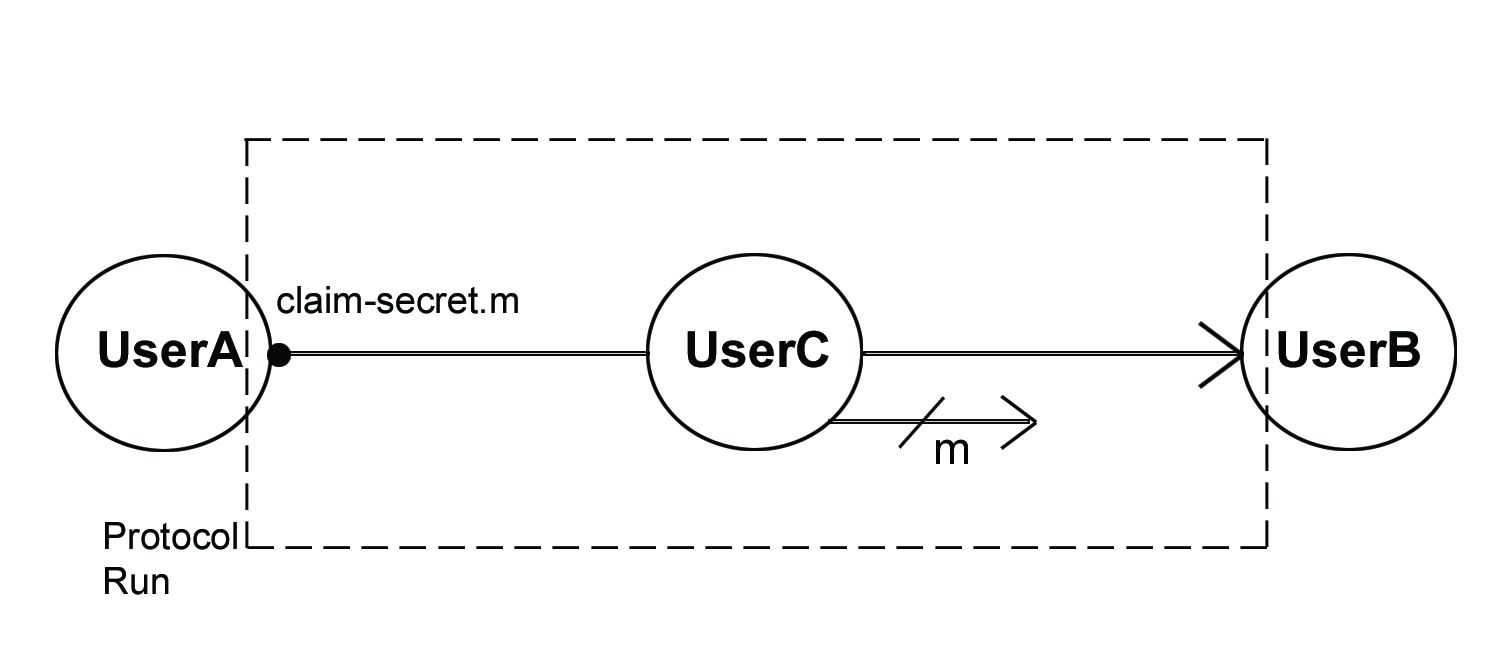}
	\caption{Model Showing a Secrecy Claim.}
	\label{secrecy}
\end{figure*}

\begin{figure*}[bpht!]
	\centering
		\includegraphics[width=8cm,height=6cm]{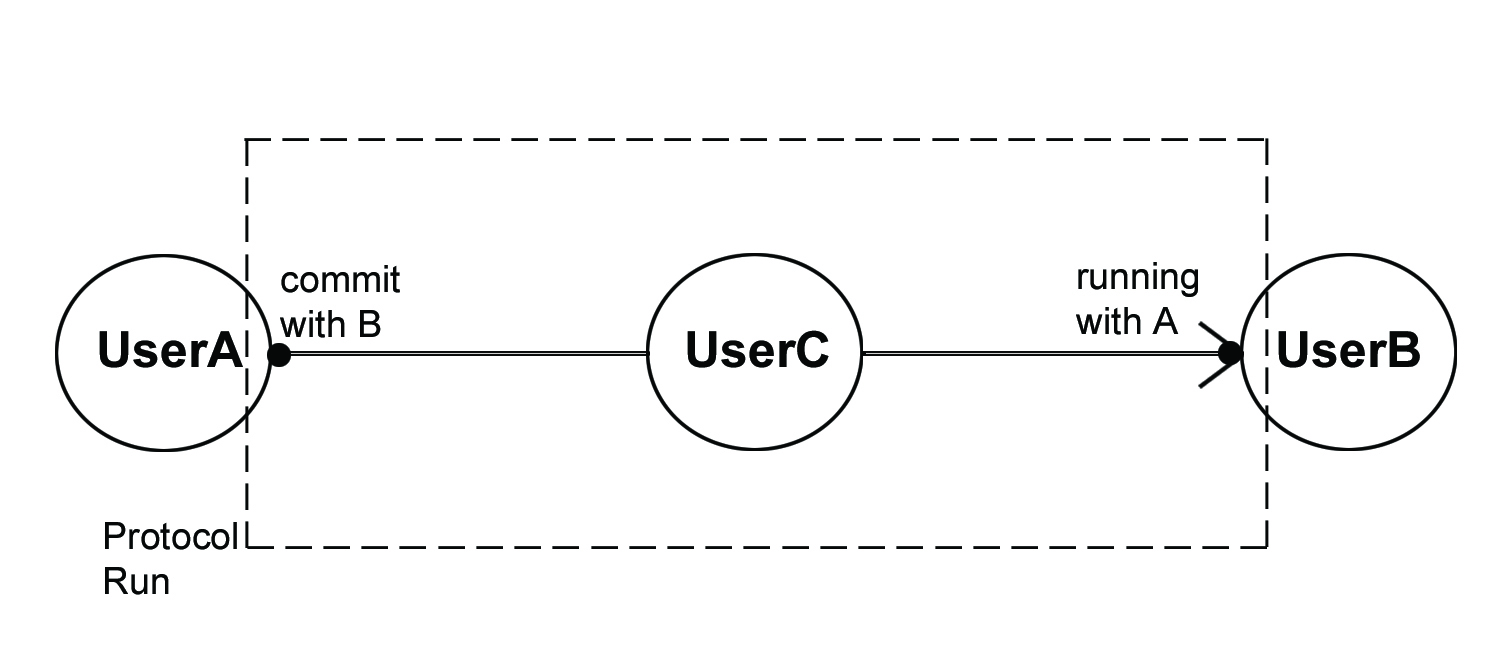}
	\caption{Model Showing an Authentication Claim.}
	\label{auth}
\end{figure*}

As per section \ref{b}, the secrecy and authentication properties can be viewed as model is shown in Fig.\ref{secrecy} and Fig.\ref{auth} respectively.

The various CTL specification for the safety properties can be written as:
\begin{itemize}
	\item AG(OldOwner(claim\_secret)) $\rightarrow$ AG(NewOwner(message))\\
	This says that the UserA (OldOwner) initiates the ownership transfer process with UserB (NewOwner). The UserA claims for the secrecy of the message. The message is kept secret during run of the protocol. Intruder cannot access any kind of details of the session.\\
	\item AG(OldOwner(Running)) $\rightarrow$ AG(NewOwner(Commit))\\
	This says that the during the run of the protocol UserA is committed to UserB. UserB is following protocol run with UserA. Any kind of information is not revealed to the intruder present in the environment.\\
	\end{itemize}
	
\section{Model Checking Results and Discussion}	
\label{f}

\begin{table}[bpht!]
\centering
\caption{Specifications of the Protocol}
	\label{tab:1}
		\begin{tabular}{|c|c|c|}
		\hline
			Sl.No. & Case & Specification\\
			\hline
			1 & UserA[1]=1 \& UserA\_claim\_secret[1][2]=1 & True\\
			\hline
			2 & UserA[1]=1 \& UserA\_commit[1][2]=1 & True\\
			\hline
			3 & UserB[2]=1 \& UserB\_running[2][1]=1 & True\\
			\hline
			4 & UserC[2]=1 \& UserC[2][1]=0 & False\\
			\hline
			5 & UserC[1]=1 \& UserC[1][1]=0 & False\\
			\hline
			\end{tabular}
	\end{table}

Table \ref{tab:1} shows the various constraints imposed on the system and results of the corresponding specifications.
\begin{enumerate}
	\item The first specification in the table depicts that UserA (Old Owner) initiates the ownership transfer process with UserB (New Owner), sends message to the CKS. The message consists of the UserA ID, Password, Nonce and OTR. The UserA claims for the secrecy of the message so that intruder present in the environment cannot access the details of the session. The secrecy property is verified through this specification. The specification simulation is shown in Fig.\ref{spec1}.	
	\begin{figure}[bpht!]
	\centering
		\includegraphics[width=8cm,height=3cm]{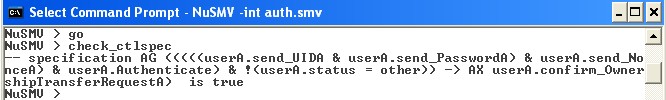}
	\caption{First Specification Verification Result Showing Result as True.}
	\label{spec1}
\end{figure}
\item The second specification in the table shows the specification as true because the UserA(old user) introduce UserB(new user) to the CKS. UserB sends ID, Nonce and Ticket to the CKS. The UserA is committed with UserB. The specification simulation is shown in Fig.\ref{spec2}.
\item The third specification in the table shows the specification as true because the UserA(old user) introduce UserB(new user) to the CKS. UserB sends ID, Nonce and Ticket to the CKS. The UserB is running with UserA. The specification simulation is shown in Fig.\ref{spec2}.

\begin{figure}[bpht!]
	\centering
		\includegraphics[width=8cm,height=3cm]{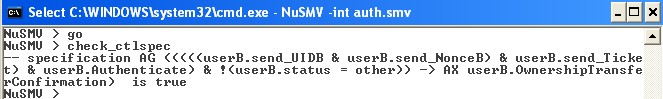}
	\caption{Second Specification Verification Result Showing Result as True.}
	\label{spec2}
\end{figure}
	 
	\item The fourth specification says that UserC (Intruder) sends credentials of UserB to the CKS, trying to access the device. This is  not possible. The specification returns false and counter example will be generated. The specification simulation is shown in Fig.\ref{spec3}.	 
	 \begin{figure}[bpht!]
	\centering
		\includegraphics[width=8cm,height=10cm]{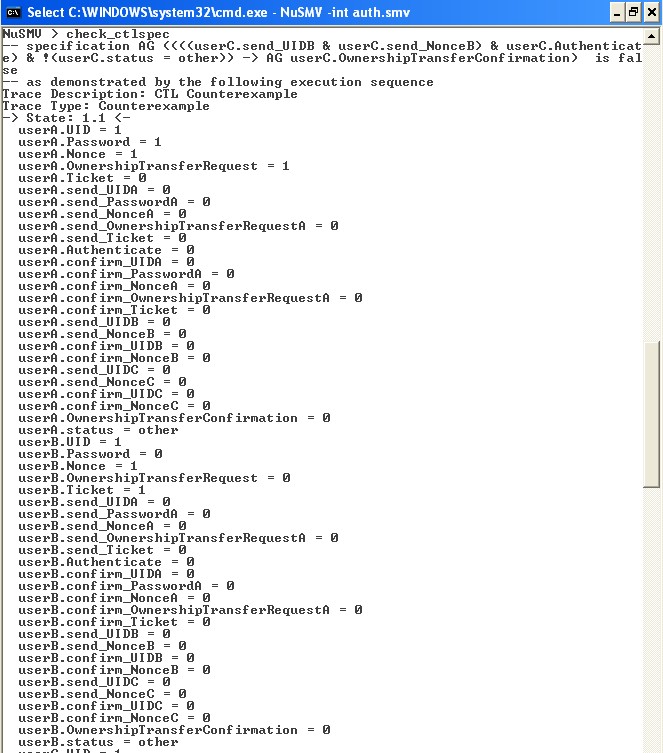}
	\caption{Fourth Specification Verification Returns False and Gives a Counterexample.}
	\label{spec3}
\end{figure} 

\item The fifth specification says that UserC (Intruder) sends UserA credentials to the CKS, trying to initiate the ownership transfer process of the device. The specification returns false and counter example will be generated. The specification simulation is shown in Fig.\ref{spec4}.	 
\begin{figure}[bpht!]
	\centering
		\includegraphics[width=8cm,height=10cm]{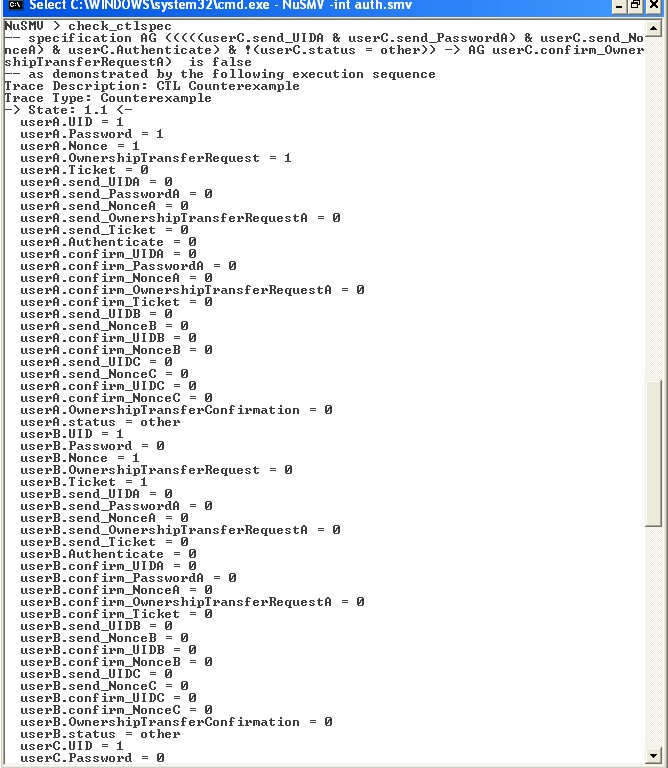}
	\caption{Fifth Specification Verification Returns False and Gives a Counterexample.}
	\label{spec4}
\end{figure} 

\end{enumerate}

\section{Conclusion}
\label{g}
In this paper, we have proposed a new ownership authentication transfer protocol for ubiquitous computing devices. The basic process of ownership authentication transfer and safety properties is described using CSP approach. We have used symbolic model checking approach to model the safety properties of the proposed protocol. The tool NuSMV helps us to verify the constraints imposed on the system by exploring the entire state space of the system. It provides a counter example along with the trace path to point to the location of error, if the system does not meet any of the constraints. The safety properties of a protocol is modeled efficiently using NuSMV. It is observed that all the constraints are met by the system developed and the proposed protocol is safe to be used in Ubiquitous environment.

\bibliographystyle{IEEEtran}
\bibliography{ref}

\end{document}